\documentclass[prl,twocolumn,superscriptaddress,showpacs]{revtex4}
\usepackage{amsmath}
\usepackage{amssymb}
\usepackage{graphicx}

\begin{document}


\title{Determination of the antiferroquadrupolar order parameters in $\mathrm{UPd_3}$}

\author{H. C. Walker}
 \author{K. A. McEwen}%
 \email{k.mcewen@ucl.ac.uk}
 \author{D. F. McMorrow}
 \affiliation{Department of Physics and Astronomy, University College
 London, Gower Street, London, WC1E 6BT, UK}%
\author{S. B. Wilkins}
\affiliation{European Synchrotron Radiation Facility, BP 200, F-38403, Grenoble Cedex, France}%

\author{F. Wastin}
\author{E. Colineau}
 \affiliation{European Commission, Joint Research Centre, Institute for Transuranium Elements,
 Postfach 2340, Karlsruhe, D-76125 Germany}
\author{D. Fort}
 \affiliation{Department of Metallurgy, University of Birmingham, Edgbaston, Birmingham, B15 2TT, UK}

\date{\today}

\begin{abstract}
By combining accurate heat capacity and X-ray resonant scattering
results we have resolved the long standing question regarding the
nature of the quadrupolar ordered phases in $\mathrm{UPd_3}$.  The
order parameter of the highest temperature quadrupolar phase has
been uniquely determined to be antiphase $Q_{\mathrm{zx}}$ in
contrast to the previous conjecture of $Q_{\mathrm{x^2-y^2}}$. The
azimuthal dependence of  the X-ray scattering intensity from the
quadrupolar superlattice reflections indicates that the lower
temperature phases are described by a superposition of order
parameters.  The heat capacity features associated with each of the
phase transitions characterise their order, which imposes
restrictions on the matrix elements of the quadrupolar operators.

\end{abstract}

\maketitle

In the past couple of decades there has been considerable interest
in the orbital ordering of $d$ and $f$ electron systems
\cite{Murakami80, Murakami81, Wilkins03, Thomas, Shiina, Aoki,
Izawa, Paixao, Wilkins, Matsumura, McMorrow}. The highly degenerate
$f$-electron shells in actinide and rare earth systems provide a
wealth of local degrees of freedom: dipolar (magnetic), quadrupolar,
octupolar etc. In localised $f$-electron systems these degrees of
freedom become potential order parameters, which can lead to
interesting and complex phase diagrams. Whilst, historically, the
order associated with magnetic dipole moments has been studied
extensively, more recently the importance of electric quadrupoles in
magnetic materials has been recognized \cite{Morin}. In classical
electrodynamics the multipole expansion suggests that the
interaction between higher order multipoles is seemingly much weaker
than between dipoles. However, the interaction between multipoles is
quantum mechanical in origin, and dipolar and quadrupolar
interactions may be equally strong. More recently, quadrupolar order
has been associated with new and interesting behaviour, such as the
novel heavy Fermion state in $\mathrm{PrFe_4P_{12}}$ \cite{Aoki},
and the exotic superconductivity in $\mathrm{PrOs_4Sb_{12}}$, which
may be mediated by quadrupolar fluctuations \cite{Izawa}.

$\mathrm{UPd_3}$ is a very interesting system, being a rare example
of a localised uranium intermetallic compound, as well as belonging
to the small class of metallic materials which exhibit long-range
quadrupolar order.  Despite intensive experimental investigation
over the past 25 years an understanding of the series of four phase
transitions below 8 K, and the exact nature of the quadrupolar
ordering in $\mathrm{UPd_3}$ have proved highly challenging. For the
first time we are now able to distinguish which order parameter is
associated with the highest temperature quadrupolar phase between
$T=7.8$ K and 6.9 K. This has been achieved using the unique
properties of X-ray resonant scattering (XRS), which may be
visualised as a process in which an incident photon promotes a core
electron to an excited intermediate state that then decays back to
the initial state, emitting a scattered photon.  XRS reveals
information about the ordering of the ground state of multipoles
associated with this intermediate state, since at resonance the
scattering length becomes a tensor \cite{Dmitri83, Dmitri84} whose
elements are directly related to the multipole moments.
Quadrupolar order describes the periodicity of charge distribution
asphericities and additional superlattice reflections are observed
in the case of antiferroquadrupolar (AFQ) ordering. Characterisation
of the dependence of such reflections on photon energy and
polarisation, and azimuthal angle (rotation around the scattering
vector) allows details of the multipolar structure to be deduced by
comparison with calculations based on tensors for the different
order parameters \cite{Paixao, Wilkins, Matsumura}.


$\mathrm{UPd_3}$ crystallizes in the double-hexagonal close-packed
(dhcp) structure (lattice parameters $a=5.73$ \AA, $c=9.66$ \AA),
 with uranium ions sitting at sites with locally hexagonal and
quasi-cubic symmetry (space group $\mathrm{P6_3/mmc}$). Four
transitions at $T_0=7.8$ K, $T_{+1}=6.9$ K, $T_{-1}=6.7$ K and
$T_2=4.4$ K have been revealed by a combination of microscopic e.g.
neutron \cite{Buyers, Steig, Walker94, McEwen98} and X-ray
\cite{McMorrow} scattering, and macroscopic measurements: ultrasound
\cite{Lingg}, heat capacity \cite{Zochowski95, Tokiwa}, magnetic
susceptibility \cite{Tokiwa, McEwen95}, thermal expansion and
magnetostriction \cite{Zochowski94}. The quadrupolar phase
transitions are very sensitive to doping Np for U\cite{HWalker}, and
Pt for Pd \cite{Zochowski94}. Neutron \cite{Steig, Walker94,
McEwen98} and X-ray \cite{McMorrow} scattering studies indicate that
the phase transition at $T_0$ is to an AFQ structure associated with
periodic lattice distortions and a doubling of the double-hexagonal
unit cell leading to superlattice reflections at
$\mathbf{Q}=(2h+1,0,l)$ in the orthorhombic notation, which will be
used hereafter.  Polarised neutron diffraction (PND) measurements
\cite{McEwen98} suggested that the phase between $T_0$ and $T_{+1}$
was $Q_{\mathrm{x^2-y^2}}$, and the earlier X-ray experiments
\cite{McMorrow} were consistent with this hypothesis. However, due
to limitations on cryostat technology, it was not possible to
perform X-ray azimuthal scans at that time in the appropriate
temperature regimes. Recently, a new crystal field model
\cite{McEwen03} with a doublet ground state on the quasi-cubic
sites, as opposed to the earlier singlet ground state model of
Buyers et al. \cite{Buyers}, provided a qualitative understanding of
the succession of quadrupolar phase transitions with suggestions for
the possible order parameters.  It also presents a framework for
understanding whether the transitions are first or second order. In
this paper, we present X-ray resonant scattering data which shows
unequivocably that below $T_0$ the AFQ structure is described by the
$Q_{\mathrm{zx}}$, rather than $Q_{\mathrm{x^2-y^2}}$, order
parameter.

Previous heat capacity experiments were made before the $T_{+1}$ and
$T_{-1}$ transitions had been identified, and so we have undertaken
new measurements to distinguish the two transitions. The heat
capacity of polycrystalline samples of $51.33$ mg $\mathrm{UPd_3}$
and $16.79$ mg $\mathrm{ThPd_3}$, used as a phonon blank, was
measured from $T=2-300$ K using a PPMS-9 Quantum Design calorimeter
in the Actinide UserLab at the Institute for Transuranium Elements.
Contributions to the heat capacity from the sample holder and grease
were measured separately and then subtracted from the total signal.

\begin{figure}[tbp]
\includegraphics[width=0.45\textwidth,bb=15 185 535 790,clip]{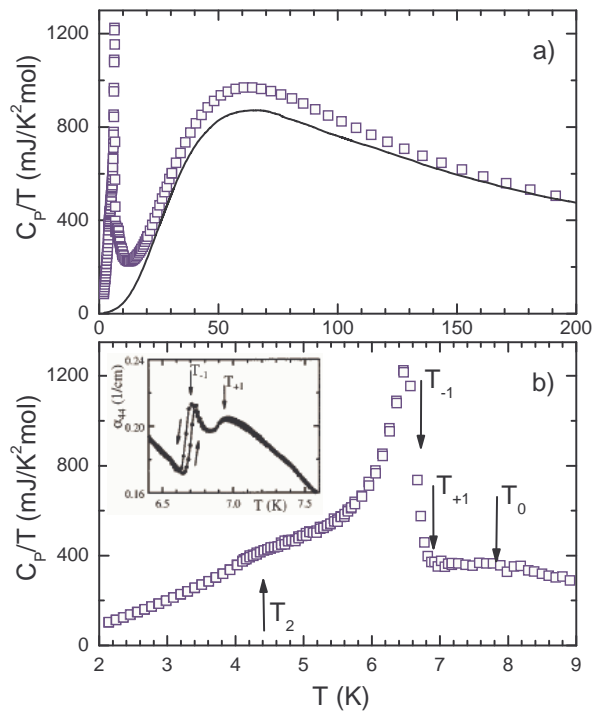}
\caption{\label{hc}a) Heat capacity of $\mathrm{UPd_3}$ and
$\mathrm{ThPd_3}$ (open blue squares and a black line respectively),
up to 200 K at which point the phonons are the dominant contribution
to the heat capacity. b) $\mathrm{UPd_3}$ heat capacity in the
region of the four transitions. The lambda anomaly clearly shows
that the first order transition is associated with $T_{-1}$ rather
than $T_{+1}$. The inset shows ultrasonics data \cite{Lingg} from
which the $T_{-1}$ and $T_{+1}$ transitions were first identified,
with hysteresis in $\alpha_{44}$ at $T_{-1}$. }
\end{figure}

Heat capacity results reveal a lambda anomaly below $T_{-1}$, which
is clearly first order, whilst there are no large entropy changes at
the other transitions, see Fig.~\ref{hc}. This affects the matrix
elements for the order parameters.  Using the three-level model
developed by McEwen et al. \cite{McEwen03}, the matrices
representing the quadrupolar operators can be written as
\vspace{-6pt}
\begin{equation}\label{eq1}
{\small\hat{Q}_{\mathrm{x^2-y^2}}=\left(\begin{array}{ccc} 0 & A & A\\
A & 0 & B\\ A & B & 0\end{array}\right)
\hat{Q}_{\mathrm{zx}}=\left(\begin{array}{ccc} 0 & A' & A'\\ A' & 0
& B'\\ A' & B' & 0\end{array}\right)}\\
\end{equation}
\vspace{-16pt}
\begin{equation*}
{\small\hat{Q}_{\mathrm{xy}}=\left(\begin{array}{ccc} 0 & -Ai & Ai\\ Ai & 0 & -Bi\\
-Ai & Bi & 0\end{array}\right)
\hat{Q}_{\mathrm{yz}}=\left(\begin{array}{ccc} 0 & A'i & -A'i\\
-A'i & 0 & B'i\\ A'i & -B'i & 0\end{array}\right)}
\end{equation*}
where the $A^{(\prime)}$ terms mix the singlet with the doublet
states, and the $B^{(\prime)}$ terms split the doublet. From Landau
theory the operators' symmetries mean that there should be at least
two first order transitions, while the data in Fig.~\ref{hc}b
indicates that there is only one strongly first order transition at
$T_{-1}$. Since the $B^{(\prime)}$ term splits the ground doublet,
leading to entropy changes at the transition, either $B$ or
$B^{\prime}$ must be $\simeq0$, to make $\hat{Q}=-\hat{Q}$ for
either $\hat{Q}_{\mathrm{x^2-y^2}}$ or $\hat{Q}_{\mathrm{zx}}$, such
that there can be just one strongly first order transition.

For the XRS experiment a single crystal of $\mathrm{UPd_3}$ was
grown, at the University of Birmingham, using the Czochralski method
with starting materials of 3N U and 4N Pd. The experiment was
performed on the ID20 beamline at the ESRF. Resonant scattering
studies can be carried out at the L and M edges of uranium
compounds, but since we wish to probe the \emph{f} electrons, we
used the $M_{\mathrm{IV}}$-edge, $E=3.726$ keV, at which dipolar
transitions connect the core $3d_{3/2}$ states to the $5f$ states.
The electric dipole transitions, E1, dominate the resonant
scattering cross section.

\begin{figure}[tbp]
\includegraphics[width=0.45\textwidth,bb=5 5 420 265,clip]{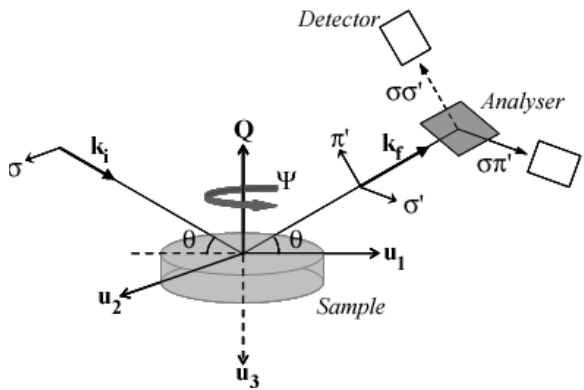}
\caption{\label{aziexpt} A schematic of the set up for measuring the
azimuthal dependence of the scattering intensity of a superlattice
reflection at \textbf{Q}. $\sigma$-polarised X-rays incident at an
angle $\theta$ on to the sample are scattered by a vector
\textbf{Q}. The scattered X-rays are separated into the rotated,
$\pi^\prime$, and unrotated, $\sigma^\prime$, channels by a
polarisation analyser before detection.  The sample stage is rotated
to give an azimuthal angle $\Psi$ about the scattering vector.  The
unit vectors $\mathbf{u}_i$ define the reference frame.}
\end{figure}

At $E=3.726$ keV only a small region of reciprocal space is
accessible. Therefore our crystal was cut with a reciprocal space
$(207)$ face such that both the $(103)$ and $(104)$ superlattice
peaks could easily be measured over a wide range of $\Psi$. The 417
mg sample was polished with $0.25$ $\mu$m diamond paste and mounted
in an azimuth displex cryostat, allowing us to operate the
diffractometer in the vertical plane, in which the incident X-rays
are $\sigma$-polarised: see Figure~\ref{aziexpt} for a schematic of
the experimental set-up.  This makes azimuthal rotation possible,
unlike in the previous horizontal geometry experiment
\cite{McMorrow}.  Polarisation analysis was performed by mounting an
Au (111) analyser crystal. Data was normalised against the monitor.

\begin{figure}[tbp]
\includegraphics[width=0.45\textwidth,bb=55 282 780 780,clip]{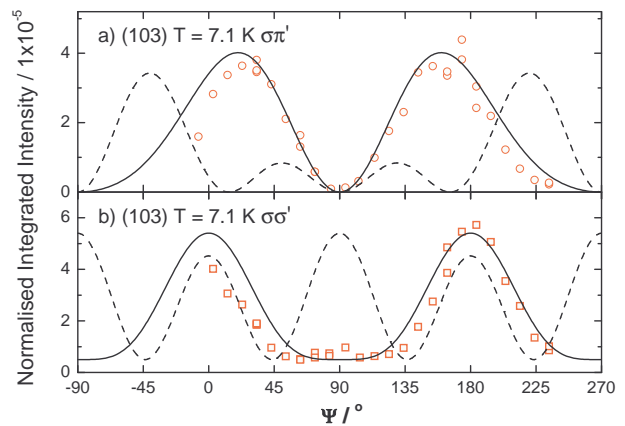}
\caption{\label{azi1} The azimuthal dependence of a) the
$\sigma\pi^\prime$ and b) the $\sigma\sigma^\prime$ scattering
intensities of the (103) peak in $\mathrm{UPd_3}$ at the U
$M_{\mathrm{IV}}$ edge, at $T=7.1$ K. Comparison is made with
calculations for the $Q_{\mathrm{zx}}$ (solid line) and the
$Q_{\mathrm{x^2-y^2}}$ (dashed line) order parameters.}
\end{figure}

The azimuthal dependence of the scattering intensity at the (103)
reflection at $T=7.1$ K, i.e. within the first quadrupolar phase, is
shown in Figure~\ref{azi1}. The azimuth angle $\Psi$ is defined
relative to the reference vector $[0\bar{1}0]$. At each azimuthal
angle, rocking curves were taken and the intensities of the
$\sigma\sigma^\prime$ and $\sigma\pi^\prime$ polarization channels
determined by fitting a Lorentzian squared lineshape minimizing
chi-squared.

The azimuthal dependence of the allowed order parameters was
calculated by summing the second rank tensors $T_n$ of the
individual uranium quadrupoles to construct the resonant scattering
length of the unit cell \cite{McMorrow}
\begin{equation}
f=\sum_nT_n\exp(i\mathbf{Q\cdot r}).
\end{equation}
The scattering amplitude is then given by
\begin{equation}
A=\mathbf{\epsilon'\cdot}f\mathbf{\cdot\epsilon}
\end{equation}
where the incident $(\epsilon)$ and scattered $(\epsilon')$
polarisations are transformed into the coordinate system of Blume
and Gibbs \cite{Blume} following the method of Wilkins et al.
\cite{Wilkins}.

\begin{figure}[tbp]
\includegraphics[width=0.45\textwidth,bb=90 300 320 450,clip]{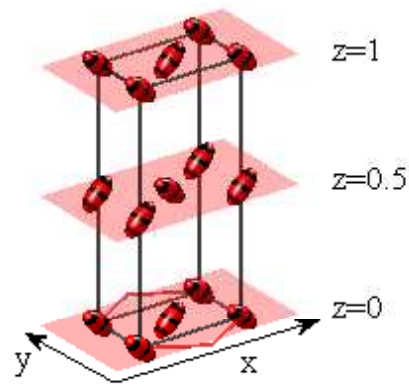}
\caption{\label{cellzx} The $Q_{\mathrm{zx}}$ AFQ structure with
antiphase stacking along the z-axis in $\mathrm{UPd_3}$ at
$T_{+1}<T<T_0$ in an orthorhombic unit cell. The U $5f$ quadrupoles
on the quasi-cubic sites are represented schematically by
ellipsoids.}
\end{figure}

Figure~\ref{azi1} shows that the $(1 0 3)$ data is in excellent
agreement with the above calculation for the azimuthal dependence of
$Q_{\mathrm{zx}}$ antiferroquadrupolar order. In the
$\sigma\pi^\prime$ channel, note the asymmetry about
$\Psi=0^{\circ}$ in both the data and $Q_{\mathrm{zx}}$ calculation,
and that the maxima are not at $\Psi=0,180^{\circ}$. In the
$\sigma\sigma^\prime$ channel, the data and calculation show a broad
minimum at $90^{\circ}$ with symmetry about $\Psi=0^{\circ}$. Since
our sample was cut with a $(207)$ reciprocal lattice face, the
scattering vector $\mathbf{Q}=(103)$ is not collinear with the face
normal (\textbf{n}). The $\sigma^\prime$ polarisation vector is
perpendicular to both \textbf{n} and \textbf{Q} so the non-collinear
nature is not observed, resulting in the symmetry about
$\Psi=0^{\circ}$ seen in $\sigma\sigma^\prime$ azimuthal
measurements. However, the $\pi^\prime$ polarisation vector lies in
the plane of \textbf{n} and \textbf{Q}, and the non-collinearity
leads to an asymmetry in the azimuthal variation in intensity.
Calculations of the $\Psi$ dependence of the scattering intensity
for the $\mathrm{Q_{x^2-y^2}}$, $\mathrm{Q_{xy}}$ and
$\mathrm{Q_{yz}}$ order parameters do not agree with the data, as
they show either the wrong periodicity, symmetry or maxima and
minima positions, and hence these order parameters can be ruled out.
$Q_{\mathrm{zx}}$ provides a natural explanation for the macroscopic
distortion to the orthorhombic cell \cite{Zochowski94} due to the
splitting of the $x-y$ symmetry, see Fig.~\ref{cellzx}. Combining
the knowledge that the transition at $T_0=7.8$ K is to a
$Q_{\mathrm{zx}}$ AFQ ordering of the $5f^2$ electrons, with the
heat capacity evidence that this transition is either second order
or very weakly first order requires $B'\simeq0$. Although the
discussion in \cite{McEwen03} assumed the order parameter was
$\mathrm{Q_{x^2-y^2}}$, with $B\simeq0$, this model is equally valid
for $\mathrm{Q_{zx}}$ as the operator matrices have the same
symmetry, see equation~\eqref{eq1}.

\begin{figure}[htp]
\includegraphics[width=0.45\textwidth,bb=50 282 780 785,clip]{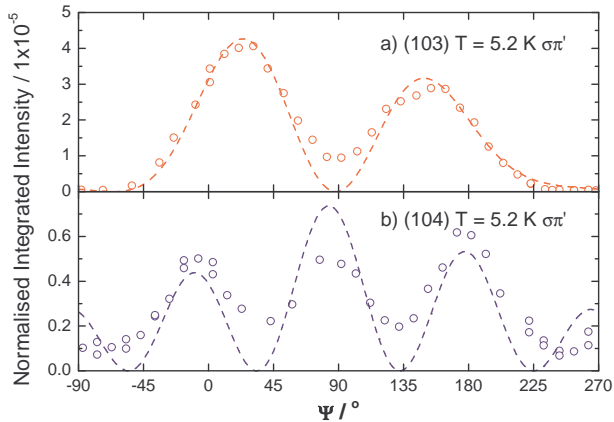}
\caption{\label{azi2} The azimuthal dependence of the
$\sigma\pi^\prime$ scattered intensity from a) the $(103)$ and b)
the $(104)$ superlattice peaks, in $\mathrm{UPd_3}$ at the U
$M_{\mathrm{IV}}$ edge, at $T=5.2$ K.  Dashed lines show least
squares fits to the data as described in the text.}
\end{figure}

In order to investigate the change in quadrupolar order in the
different phases of $\mathrm{UPd_3}$ seen in various measurements
and the temperature dependence of the RXS \cite{McMorrow}, azimuthal
measurements were made about both $\mathbf{Q}=(103)$ and $(104)$ at
$T=5.2$ K to study the third quadrupolar phase, see Fig.~\ref{azi2}.
These measurements were made in the $\sigma\pi^\prime$ channel only,
since there is considerable charge scattering interference in the
$\sigma\sigma^\prime$ channel,
From the first order lambda peak in the heat capacity at $T_{-1}$,
we expect to find evidence of an admixture of $Q_{\mathrm{x^2-y^2}}$
in this phase. It is very difficult to stabilize the cryostat
reliably between 6.7 and 6.9 K for extended periods and so we have
concentrated on measuring azimuthal scans in the third quadrupolar
phase. However, the heat capacity data indicates that at $T_{+1}$,
$\Delta S\sim0$, which would be consistent with the evolution of
$Q_{\mathrm{yz}}$ order.

The azimuthal dependence of scattering from the (103) peak reflects
the part of the AFQ structure which varies in antiphase along the
c-axis, whilst that for the (104) peak reflects that which is
uniform along the c-axis (in-phase stacking).  Clearly the azimuthal
dependence of the $\mathbf{Q}=(103)$ scattering at 5.2 K is more
complicated than in the higher temperature $Q_{\mathrm{zx}}$ phase.
A least squares fit to the data, varying the contributions from the
allowed quadrupolar moments, indicates that $Q_{\mathrm{zx}}$ still
predominates in the order parameter, but that $Q_{\mathrm{xy}}$ and
$Q_{\mathrm{x^2-y^2}}$ are also present.  The fit in Figure
~\ref{azi2}a) is derived from an anomalous scattering tensor in
which $Q_{\mathrm{zx}}$, $Q_{\mathrm{xy}}$ and
$Q_{\mathrm{x^2-y^2}}$ are present in the ratio 77:10:13. The
$(104)$ reflection, which is observed below $T_{+1}$, indicates the
existence of in-phase (along the c-axis) components to the order
parameter, as well as the antiphase components deduced from the
(103) reflection. The fit to our data in Figure ~\ref{azi2}b), which
reproduces the periodicity of the $(104)$ data, indicates that the
most significant components are $Q_{\mathrm{xy}}$ and
$Q_{\mathrm{yz}}$ in the ratio 2:1.

In conclusion, our new heat capacity measurements reveal only one
strong first-order transition at $T=T_{-1}$, whilst $\Delta
S\simeq0$ at both $T_{+1}$ and $T_0$. These observations constrain
the quadrupolar operator matrix elements within our model
\cite{McEwen03}. Our azimuthal XRS experiments have unambiguously
identified the order parameter for the first antiferroquadrupolar
phase, $T_{+1}<T<T_0$, in $\mathrm{UPd_3}$ as $Q_{\mathrm{zx}}$.
This may be reconciled with the PND $Q_{\mathrm{x^2-y^2}}$ result,
since those measurements were necessarily made in a magnetic field.
Calculations show that the wavefunctions and energies of the crystal
field states in our model \cite{McEwen03} are significantly modified
in fields of a few Tesla. We will examine with XRS the field
behaviour of the different quadrupolar phases in future experiments.
Our results have provided further valuable insight into the nature
of the other quadrupolar phases, indicating a surprisingly complex
sequence of order parameters leading to a rotation of the charge
densities.

\begin{acknowledgments}
We thank the European Community-Access to Research Infrastructures
action of the Improving Human Potential Programme (IHP), contract
HPRI-CT-2001-00118. H.C.W. thanks EPSRC for financial support, and
D.F.M. thanks the Royal Society for a Wolfson Research Merit Award.
The authors thank L. Paolasini, C. Detlefs and P. Deen for useful
discussions and their assistance.
\end{acknowledgments}

\bibliography{paper}

\end{document}